\documentclass{emulateapj}
\usepackage{amsmath}
\usepackage{amsfonts}

\shorttitle{Evolution of EW(Ha)}
\shortauthors{Mattia Fumagalli et al.}
\slugcomment{Accepted for publication in The Astrophysical Journal Letters}

\begin{document}
\title{H$\alpha$ Equivalent Widths from the 3D-HST survey: evolution with redshift and dependence on stellar mass \footnote{B\lowercase{ased on observations made with the} NASA/ESA H\lowercase{ubble} S\lowercase{pace} T\lowercase{elescope, obtained at the} S\lowercase{pace} T\lowercase{elescope} S\lowercase{cience} I\lowercase{nstitute, which is operated by the }A\lowercase{ssociation of }U\lowercase{niversities for} R\lowercase{esearch in} A\lowercase{stronomy}, I\lowercase{nc., under} NASA \lowercase{contract} NAS 5-26555. T\lowercase{hese observations are associated with programs 12177, 12328}}}

\author{Mattia Fumagalli\altaffilmark{1}, Shannon G. Patel\altaffilmark{1}, Marijn Franx\altaffilmark{1}, Gabriel Brammer\altaffilmark{2}, Pieter van Dokkum\altaffilmark{3}, Elisabete da Cunha\altaffilmark{5}, Mariska Kriek\altaffilmark{4}, Britt Lundgren\altaffilmark{3}, Ivelina Momcheva \altaffilmark{3}, Hans-Walter Rix\altaffilmark{5}, Kasper B. Schmidt\altaffilmark{5}, Rosalind E. Skelton\altaffilmark{3}, Katherine E. Whitaker\altaffilmark{3}, Ivo Labbe\altaffilmark{1},  Erica Nelson\altaffilmark{3}}

\altaffiltext{1}{Leiden Observatory, Leiden University, P.O. Box 9513, 2300 RA Leiden, Netherlands}

\altaffiltext{2}{European Southern Observatory, Alonso de Córdova 3107,Casilla 19001, Vitacura, Santiago, Chile}

\altaffiltext{3}{Department of Astronomy, Yale University, New Haven, CT 06511, USA}

\altaffiltext{4}{Department of Astronomy, University of California, Berkeley, CA 94720, USA}

\altaffiltext{5}{Max Planck Institute for Astronomy (MPIA), Königstuhl 17, 69117 Heidelberg, Germany }

\begin{abstract}

We investigate the evolution of the H$\alpha$ equivalent width, EW(H$\alpha$), with redshift and its
dependence on stellar mass, using the first data from the 3D-HST survey, 
a large spectroscopic Treasury program with the HST-WFC3. 
Combining our H$\alpha$ measurements of 854 galaxies at $0.8<z<1.5$ with those of ground based surveys at lower and higher redshift, we can consistently determine the evolution of the EW(H$\alpha$) distribution from z=0 to z=2.2.
We find that at all masses the characteristic EW(H$\alpha$) is decreasing towards the present epoch, and that at each redshift the EW(H$\alpha$) is lower for high-mass galaxies.
We find EW(H$\alpha$) $\sim (1+z)^{1.8}$ with little mass dependence.
Qualitatively, this measurement is a model-independent confirmation of the evolution of star forming galaxies with redshift.
A quantitative conversion of EW(H$\alpha$) to sSFR (specific star-formation rate) is model dependent,
because of differential reddening corrections between the continuum and the Balmer lines.
The observed EW(H$\alpha$) can be reproduced with the characteristic evolutionary history for galaxies, whose star formation rises with cosmic time to $z \sim 2.5$ and then decreases to $z$ = 0. This implies that EW(H$\alpha$) rises to 400 $\rm\AA$ at $z = 8$. The sSFR evolves faster than EW(H$\alpha$), as the mass-to-light ratio also evolves with redshift. We find that the sSFR evolves as $(1+z)^{3.2}$, nearly independent of mass, consistent with previous reddening insensitive estimates.
We confirm previous results that the observed slope of the sSFR-$z$ relation is steeper than
the one predicted by models, but models and observations agree in finding little mass dependence.

\end{abstract}

\keywords{galaxies: evolution — galaxies: formation — galaxies: high-redshift}

\section{Introduction}

Several studies have combined different star formation indicators in order to study the evolution of star-forming galaxies (SFGs) with redshift.
At a given redshift low mass galaxies typically form more stars per unit mass (i.e., specific star-formation rate, sSFR) than more massive galaxies
(Juneau et al. 2005, Zheng et al. 2007, Damen et al. 2009). 
In addition the sSFR of galaxies with the same mass increases at higher redshift. 
However, semi-analytical models and observations are at odds with regards to the rate of decline of the sSFR towards low-redshift (Damen et al. 2009, Guo et al. 2010).

One of the main observational caveats is that
most of the studies covering a wide redshift range use
diverse SFR indicators (such as UV, IR, [OII], H$\alpha$, SED fitting). This is a consequence of the fact that it is difficult to use 
the same indicator over a wide range of redshifts.  
One therefore has to rely on various conversion factors, often intercalibrated at $z=0$ and 
re-applied at higher redshift.
\begin{figure*}[!t!]
\centering
\includegraphics[width=18cm]{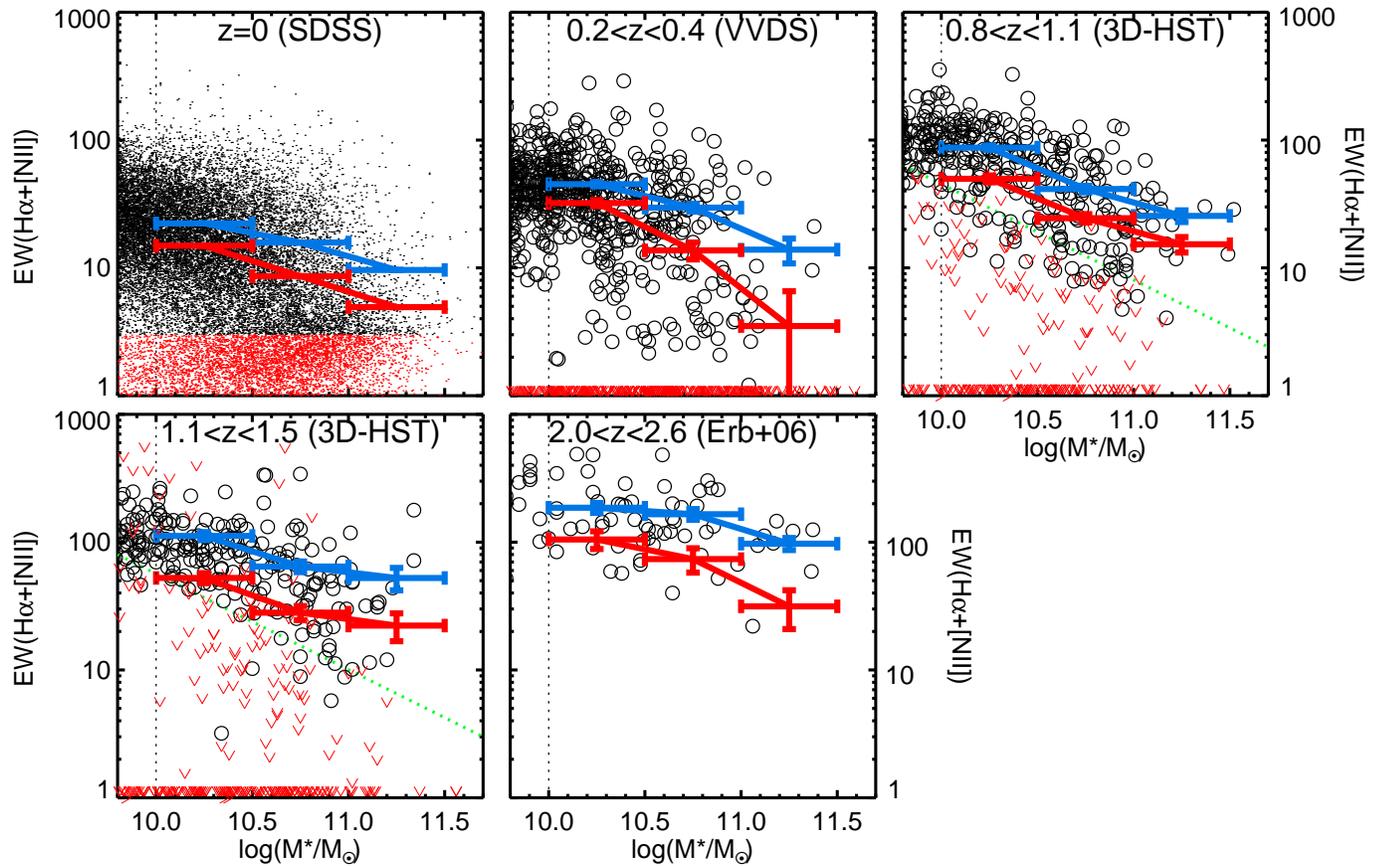}											     	  

\caption{EW(H$\alpha$) against mass, for different redshift samples. Vertical lines represent the limiting mass of the analysis. Black symbols are objects with H$\alpha$ detection with S/N $>$ 3 and red arrows represent upper limits. The green diagonal lines represent the detection limit of the 3D-HST data. Blue solid lines represent the mean EW(H$\alpha$) of detected SFGs, in 0.5 dex mass bins. Red solid lines represent the mean EW(H$\alpha$) of all galaxies, assuming EW(H$\alpha$)=0 for non-detected objects. Errors of the means are computed with a bootstrap approach. At each redshift higher mass galaxies have lower EW(H$\alpha$) than less massive objects.}
\label{EWmass}

\end{figure*}

A well-calibrated standard indicator of the SFR is the H$\alpha$ luminosity (Kennicutt, 1998). 
However, H$\alpha$  is shifted into the infrared at $z > 0.5$, and it is difficult to measure due to the limitations of ground-based near-IR spectroscopy.
Comparing measures of H$\alpha$ at different redshifts has therefore been a challenge. Most of the H$\alpha$ studies at high redshift are based on narrow-band photometry (e.g. the HiZELS survey, Geach et al. 2008).

The 3D-HST survey (Brammer et al., 2012) 
provides a large sample of rest-frame optical spectra with the WFC3 grism,
which includes the H$\alpha$ emission in the redshift range $0.8 < z < 1.5$. Taking advantage of the first data from 
the survey (45\% of the final survey products) we investigate for the first time 
the star formation history (SFH) of the Universe with H$\alpha$ spectroscopy, using a consistent SFR indicator over a wide redshift range. 

We evaluate the dependence of the H$\alpha$ Equivalent Width, EW(H$\alpha$), on stellar mass ($\rm M_{*}$) and redshift (up to $z \sim 2$), 
comparing the 3D-HST data with other surveys
in mass selected samples with $\rm M_{*} > 10^{10} M_{\odot}$. Since EW(H$\alpha$) is defined as the ratio of the H$\alpha$ 
luminosity to the underlying stellar continuum, it represents a measure of the 
the current to past average star formation. It is therefore a model independent, directly observed proxy for sSFR. 

We also derive SFRs from the H$\alpha$ fluxes.
We evaluate the mean sSFR in stellar mass bins and study its evolution with redshift.
The slope of the sSFR-$z$ relation in different mass bins indicates how fast the star formation is quenched in galaxies of various masses. 
Finally, we compare our findings to other studies
(both observations and models), discussing the physical implications and reasons for any disagreements.

\begin{figure*}
\includegraphics[width=18cm]{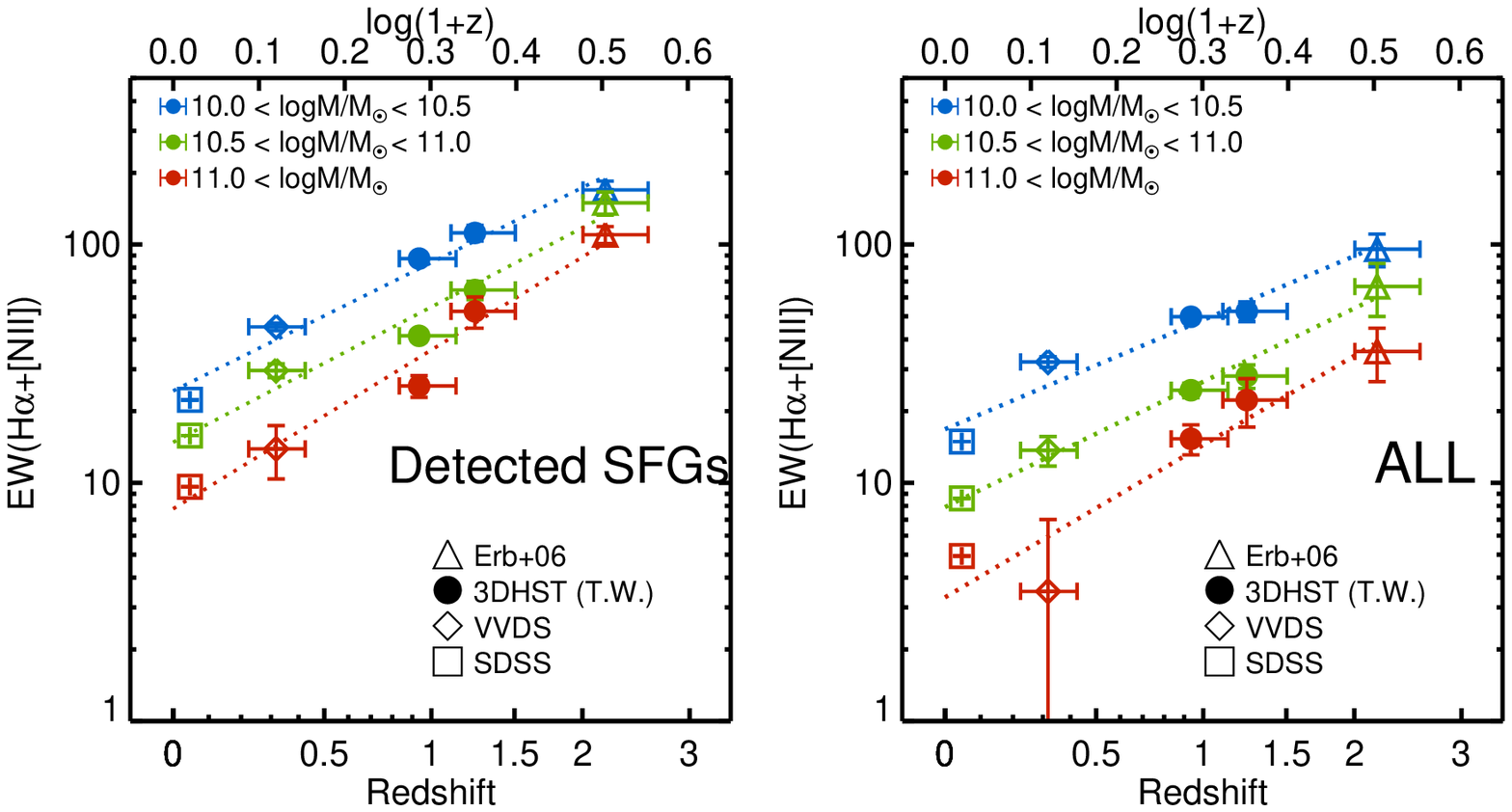} 
										     	  \\
\includegraphics[width=18cm]{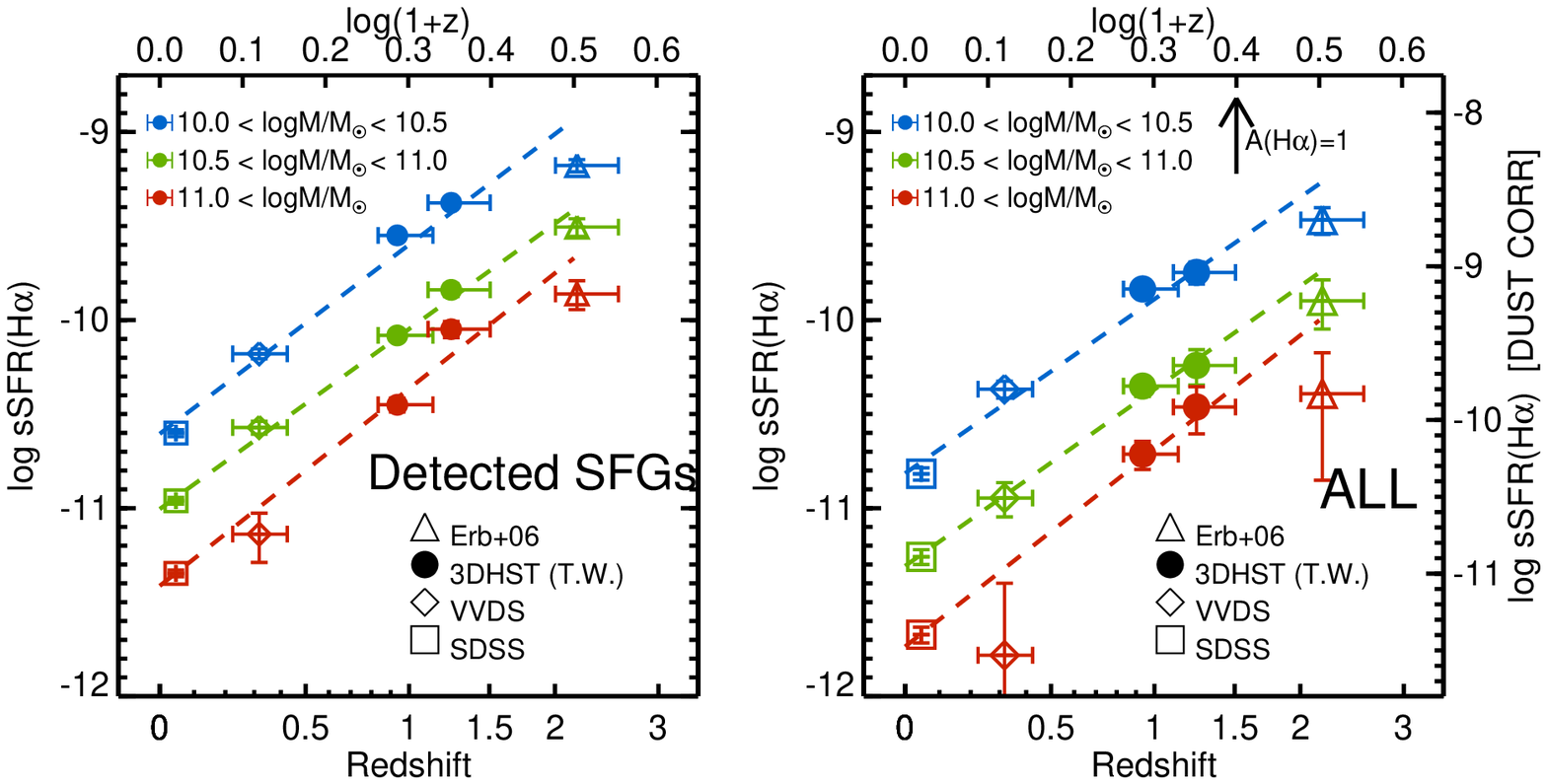}
\caption{Evolution of EW(H$\alpha$) (top) and sSFR(H$\alpha$) (bottom) with redshift, in different mass bins, for SFGs (left) and all objects (right). Errors on the average EWs have been evaluating through bootstrapping. Dotted lines are the best fit power laws 
${\rm EW}(z) \sim  (1+z)^{p}$. At fixed mass the average EW(H$\alpha$) and sSFR(H$\alpha$) increase with redshift, with a power law of ${\rm EW(H\alpha)} \sim (1+z)^{1.8}$ and ${\rm sSFR(H\alpha)} \sim (1+z)^{3.3}$ with little mass dependence. The effect of a luminosity dependent dust correction (Garn et al. 2010) correction is shown by the right axis. The effect of A(H$\alpha$)=1 is shown by the black arrow.}
\label{evolutionEW}
\end{figure*}

\section{Data}

\subsection{3D-HST}
We select the sample from the first available 3D-HST data. They include 25 pointings in the COSMOS field, 6 in GOODS-South, 12 in AEGIS, 28 in GOODS-North\footnote{From program GO-11600 (PI:
B. Weiner)}. Spectra have been extracted with the aXe code (Kummel et al., 2009). 
Redshifts have been measured via the combined photometric and  spectroscopic information using a modified version of the EAZY code (Brammer, van Dokkum, Coppi, 2008), as shown in Brammer et al. (2012).  Stellar masses were determined using the FAST code by Kriek et al. (2009), using Bruzual \& Charlot (2003) models and assuming a Chabrier (2003) IMF. The FAST fitting procedure relies on photometry from the NMBS catalogue 
(Whitaker et al. 2011) for the COSMOS and AEGIS fields, the MODS catalog for the GOODS-N (Kajisawa et al. 2009) and the FIREWORKS catalogue for GOODS-S (Wuyts et al. 2008).

The mass completeness limit is $\rm log(M_{*}/M_{\odot}) > 10$ at $z$=1.5 (Wake et al. 2011, Kajisawa et al. 2010). 
We select objects with $0.8<z<1.5$; this resulted in a sample of 2121 galaxies with $\rm log(M_{*}/M_{\odot}) > 10$.
 
In slitless spectroscopy, spectra can be contaminated by overlapping 
spectra of neighboring galaxies. The aXe package provides a quantitative estimate of the contamination 
as a function of wavelength, which can be subtracted from the spectra. We conservatively use spectra where the average contribution of contaminants is lower than 10\% and for which more than 75\% of the spectrum falls on the detector. 
After this selection we have 854 objects in the redshift range $0.8<z<1.5$ (40\% of the objects). The final sample is not biased with respect to the mass relative to the parent sample.

Line fluxes and EWs were measured as follows.\footnote{Through the entire paper the quoted EWs are rest-frame values.}
We fit the 1D spectra with a gaussian profile plus a linear continuum 
in the region where H$\alpha$ is expected to lie. 
We subtract the continuum from the fit and measure the residual 
flux within 3$\sigma$ from the line center of the gaussian. 
Errors are evaluated including the contribution from the error on the continuum. 
We distinguish between detections and non-detections of H$\alpha$ with a S/N threshold of 3.
The typical 3$\sigma$ detection limit corresponds to $\rm SFR = 2.8 M_{\odot}  yr^{-1}$ at $z$=1.5 (Equation 2). 

Due to the low resolution of the WFC3 grism, the H$\alpha$ 
and [NII] lines are blended.
In this work EW(H$\alpha$) therefore includes the contribution from [NII].
For the other datasets, which have higher spectral resolution, we combine H$\alpha$ and [NII] for consistency with 3D-HST. 

\subsection{SDSS}
We retrieve masses and EW(H$\alpha$) for the SDSS galaxies from the MPA-JHU catalogue of the SDSS-DR7. 
Masses are computed based on fits to the photometry, following Kauffmann et al. (2003) and Salim et al. (2007).
At redshift $0.03 < z < 0.06$, for masses higher than $M_{*} = 10^{10} M_{\odot}$, the SDSS 
sample is spectroscopically complete in stellar mass (Jarle Brinchmann, private communication). 
We consider as detections only measurements greater than 3$\rm\AA$, as the ones with $\rm  EW < 3 \AA$ are affected by uncertainties in the stellar continuum subtraction (Jarle Brinchmann, private communication).

In the redshift range $0.03 < z < 0.06$, the spectroscopic fiber of SDSS does not cover the entirety of most galaxies. 
As a consequence sSFR are evaluated with emission line fluxes and masses from the fiber alone.

\subsection{VVDS}
The VIMOS VLT Deep Survey (VVDS, Le Fèvre et al. 2005) is a wide optical selected survey of 
distant galaxies. H$\alpha$ is covered by the VIMOS spectrograph at $0.0 < z < 0.4$.

Lamareille et al. (2009) released a catalog of 20,000 galaxies with line measurements, complete down to $\rm M_{*} = 10^{9.5} M_{\odot}$ at $z = 0.5$ . Masses are retrieved from the VVDS catalog; they have been computed though a Bayesian approach based on photometry (equivalent to Kauffmann et al. 2003 and Tremonti et al. 2004), and are relative to a Chabrier IMF.
We select a sample with redshift $0.2 < z < 0.4$ and $\rm  M_{*} > 10^{10.0} M_{\odot}$ containing 741 objects, of which 477 (64 \%) have an H$\alpha$ measurement with S/N $> 3$. The percentage of H$\alpha$ detected objects drops to 32\% at masses $\rm M_{*} > 10^{11.0} M_{\odot}$.

\subsection{High redshift data}
\label{High redshift data}
Erb et al. (2006) published EW(H$\alpha$) for galaxies selected with the BX criterion (Steidel et al. 2004), 
targeting SFGs at redshift $2.0 < z < 2.5$. We evaluate the completeness of the sample as follows. From the FIREWORKS catalogue (Wuyts et al. 2008) we reconstruct the BX selection
and evaluate the fraction of objects with spectroscopically confirmed redshift $2.0 < z < 2.5$ that fall 
in the BX selection. Percentages are $44 \%$, $32 \%$ and $27 \%$ for mass limited samples with $\rm log(M_{*}/M_{\odot}))$ = 10.0-10.5, $\rm  log(M_{*}/M_{\odot}) =$ 10.5-11, $\rm log(M_{*}/M_{\odot}) > 11.0$.

\section{The EW(H$\alpha$) - mass relation}
\label{The EW(Ha) - mass relation}
We first study how EW(H$\alpha$) depends on stellar mass in each available data set.
The 3D-HST sample has been divided in two redshift bins,
$0.8 < z < 1.1$ and $1.1 < z < 1.5$. The results are shown in Figure \ref{EWmass}. At each redshift, highest mass galaxies have lower EW(H$\alpha$). Note however, that there is a large scatter in the relation. 
We quantify the trend in the following way: we determine the average EW(H$\alpha$) in three 0.5 dex wide mass bins ($\rm 10.0 < log(M_{*}/M_{\odot} < 10.5$, 
$\rm  10.5<log(M_{*}/M_{\odot})<11$, $\rm log(M_{*}/M_{\odot}) > 11.0$) and evaluated its error through bootstrapping the sample. The mean EW(H$\alpha$) in a given mass bin is obtained in two ways: (1) using only detected, highly SFGs (blue lines in Figure \ref{EWmass}) and (2) using all galaxies but
assigning EW(H$\alpha$)=0 to the objects detected in H$\alpha$ with S/N $<$ 3 (red lines in Figure \ref{EWmass}).
For the $z=2.2$ data, we use the the FIREWORKS catalog to establish the fraction of galaxies excluded by the BX selection and therefore give an estimation of EW(H$\alpha$) for all galaxies.

Using either method we find that an EW(H$\alpha$)-mass relation is in place at each redshift, not just for strongly star forming objects but also for the entire galaxy population. Galaxies in the lowest mass bin ($\rm 10.0 < log(M_{*}/M_{\odot}) < 10.5 $) have on average an EW(H$\alpha$) which is 5 times higher than galaxies in the highest mass bin ($\rm log(M_{*}/M_{\odot}) > 11.0$). 

We discuss the evolution of the EW(H$\alpha$)-mass relation with redshift in the next section of the Letter. However, it is immediately evident from Figure \ref{EWmass} 
that the 3D-HST survey targets galaxies with EW(H$\alpha$) typical of an intermediate regime between what is seen at z=0 and what is seen at higher redshift.  
In other words the EW(H$\alpha$)-mass relation seems to rigidly shift towards higher EW(H$\alpha$) at higher redshifts.

A considerable fraction of detected galaxies in 3D-HST have $\rm EW(H\alpha) > 30 \AA$, 
while in SDSS similar objects are extremely rare: 3.8\%, 1.4\% and 0.4\% for increasing mass samples.
This study can be seen as an extension of the findings of van Dokkum et al. (2011), who reported 
that massive galaxies at $z>1$ show a wider range of EW(H$\alpha$) compared to galaxies in the local Universe.
Following this trend with redshift, in the $z>2$ bin we find typical EW(H$\alpha$) of 150 $\rm\AA$ for SFGs. 
Such high values of EW(H$\alpha$) represent just 11\% of the 3D-HST sample 
(14.6\%, 8.1\% and 4.7\% respectively for increasing mass bins).

\section{The Evolution of EW(H$\alpha$) with redshift}

The evolution of EW(H$\alpha$) with redshift can be seen as an observational (i.e. model-independent) proxy for the sSFR-$z$ relation.
Figure \ref{evolutionEW} (top panels) shows the redshift evolution of the average EW(H$\alpha$)
in different mass bins for the detected SFGs 
(top left panel) and for all galaxies (top right panel).
A substantial increase of the EW(H$\alpha$) is seen at higher redshifts in both samples. 
We therefore infer that evolution of the SFR is not a byproduct of selection effects from different SFR indicators. 
At $0.8<z<1.5$, a galaxy has on average an EW(H$\alpha$) that is 3-4 times higher than that of an object of comparable mass in the local universe. 
For each mass bin we parametrize the redshift evolution of the EW(H$\alpha$) as follows:
\begin{equation}
{\rm EW(H\alpha)}(z) \sim {\rm A} \times (1+z)^{p}
\end{equation}
The coefficient $p$ has an average value of $1.8$, with little dependence on mass (best fit values are listed in Table \ref{Slopes}). As can be deduced from Table 1 there may be a weak mass dependence such that the relations steepen with mass; however, the difference between the slopes at the lowest and highest mass bin is not statistically significant. 

This indicates that the decrease of EW(H$\alpha$) happens at the same rate for all galaxies irrespectively of their masses. As seen in the right panel of Figure 2, the addition of non-SFGs amounts to a negative vertical shift in the EW(H$\alpha$) but not to a change in the slope of the relation.

An uncertainty is the effect of dust on the EW(H$\alpha$). Without more measurements we cannot state what effect dust has, and in literature there is disagreement on the relative extinction suffered by the nebular emission lines and the underlying stellar continuum (Calzetti et al. 2000, Erb et al. 2006, Wuyts et al. 2011). However, the data motivated model described in Section 6 suggests that dust has a mild effect on the EW(H$\alpha$).

\section{The sSFR(H$\alpha$) - mass relation and its evolution with redshift}
\label{The sSFR(Ha) - mass relation}

\begin{figure}
\centering
\includegraphics[width=8cm]{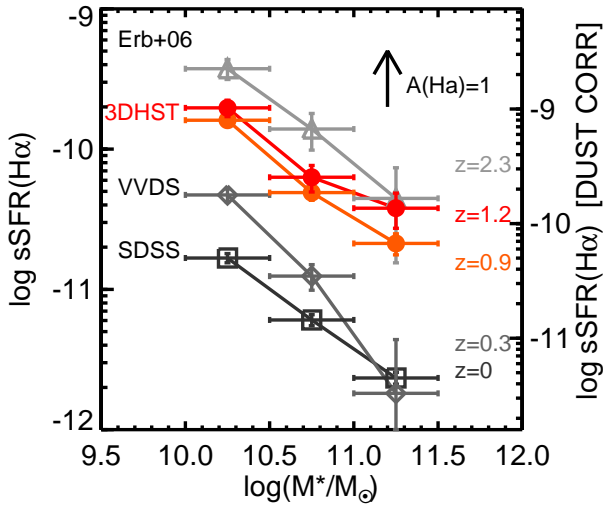}
\caption{Mean values of sSFR(H$\alpha$) in 0.5 dex mass bins at various redshift for SDSS, VVDS, 3D-HST and from Erb et al. (2006). At each redshift more massive galaxies have less sSFR(H$\alpha$) than less massive ones. The effect of a luminosity dependent dust correction (Garn et al. 2010) correction is shown by the right axis. The effect of A(H$\alpha$)=1 is shown by the black arrow.}

\label{sSFRmass}
\end{figure}

The EW(H$\alpha$) has the advantage that it is a direct observable,
but it is more difficult to interpret than a more physical quantity like
the specific star formation rate. The latter can only be derived with a
proper extinction correction for H$\alpha$. We lack this information, as we
do not have a proper Balmer decrement measurement. In the following the
briefly explore the specific star formation rate (sSFR) evolution implied by assuming
no extinction and later discuss the effect of a dust correction to the measured slopes of the sSFR-$z$ relation.

The SFR is derived from the H$\alpha$ flux\footnote{We assume a [NII]/(H$\alpha$+[NII]) ratio of 0.25 for the 3D-HST sources.} using Kennicutt (1998):
\begin{equation}
\rm SFR(H\alpha)[M_{\odot} yr^{-1}] = 7.9 \times 10^{-42} \times L(H\alpha)[erg/s] \times 10^{-0.24}
\end{equation}
where the $10^{-0.24}$ factor accounts for a conversion to the Chabrier IMF, from Salpeter (as in Muzzin et al. 2010). 

Figure \ref{sSFRmass} shows the mean value of sSFR(H$\alpha$) in different stellar mass bins, at different redshifts. In each redshift bin higher mass galaxies have lower sSFR(H$\alpha$). 

In Figure \ref{evolutionEW} (bottom panels) we show the redshift evolution of the average sSFR(H$\alpha$)
in different mass bins for detected SFGs 
(bottom left panel) and for all galaxies (bottom right panel).
A typical galaxy at z=1.5 has as sSFR(H$\alpha$) 15-20 times higher 
than a galaxy of the same mass at z=0. In each mass bin we fit the evolution of the sSFR in redshift with a power law:
\begin{equation}
{\rm sSFR} (z) \sim {\rm B_{M}} \times (1+z)^{n}
\end{equation}
obtaining a value of $n = 3.2 \pm 0.1$. As can be deduced from Table 1 there may be a weak mass dependence such that the relations steepen with mass; however, the slopes at the lowest and highest mass bin are not statistically different.

The sSFR-$z$ relation is steeper than the EW-$z$ relation, because of the additional evolution of the M/L ratio:
\begin{equation}
\rm EW/sSFR \sim  L(H\alpha)/L_{R} \times M_{*}/ (K*L(H\alpha)) \sim M/L_{R}
\end{equation}
where K is the conversion factor in Equation 2 and $\rm L_{R}$ is the R-band luminosity.

Implementing a luminosity dependent dust correction for H$\alpha$ (Garn et al. 2010, shown with the right axis in Figure \ref{evolutionEW}, bottom right) would increase the value of $n$ to $3.7\pm 0.1$. However, several studies (Sobral et al. 2012, Dominguez et al. 2012, Momcheva et al., submitted) have indicated that a better indicator for the H$\alpha$ extinction
at different redshifts is the stellar mass, and that the H$\alpha$ extinction depends strongly on mass but little on redshift (at constant mass). The Garn \& Best 2010 relation gives median A(H$\alpha$) of 1, 1.5 and 1.7 mag for the increasing mass bins in this study. A mass-dependent dust correction impacts the normalization of the sSFR but not the slope $n$. 

Our implied evolution of the sSFR compares
well to results from literature. For example, Damen et al. (2009) found
$n=4 \pm 1$ based on UV + IR inferred sSFRs, and Karim et al. (2011) found
$n=3.50 \pm 0.02$ for SFGs and $n=4.29 \pm 0.03$ for all galaxies, based on stacked radio imaging.

All results indicate an evolution which is steeper with redshift than
semianalytical models (Guo \& White 2008, Guo et al. 2011), who
find slopes close to $n$=2.5.
All studies find that the slope does not depend on the stellar mass out so $z$=2.
In short, our results are consistent with previous determinations.

\section{Linking the characteristic SFH of galaxies and EW(H$\alpha$)}

\begin{figure*}
\vspace{0.5cm}
\centering
 \includegraphics[height=7.5cm]{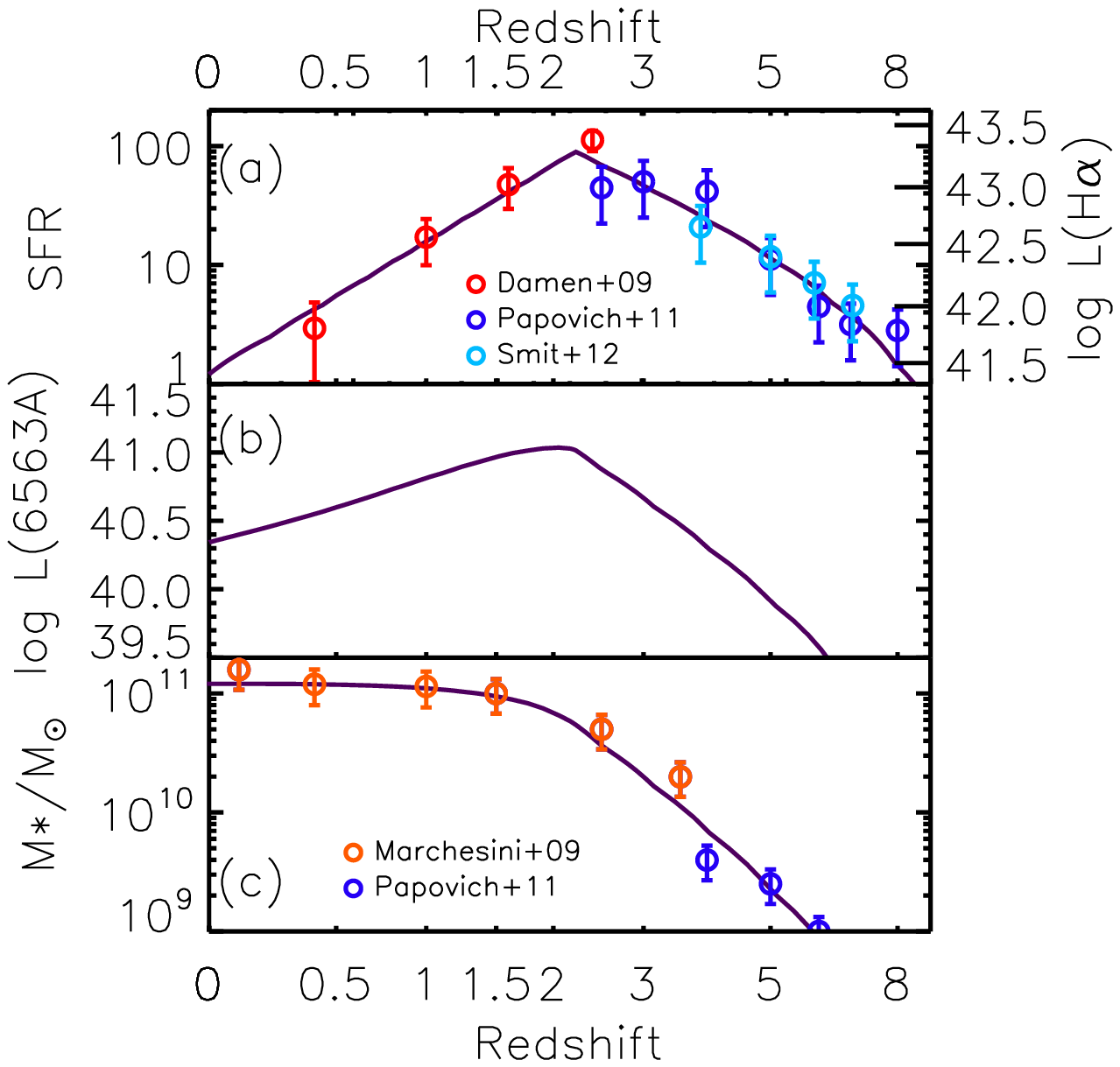} 
\hspace{0.2cm}
\includegraphics[height=7.5cm]{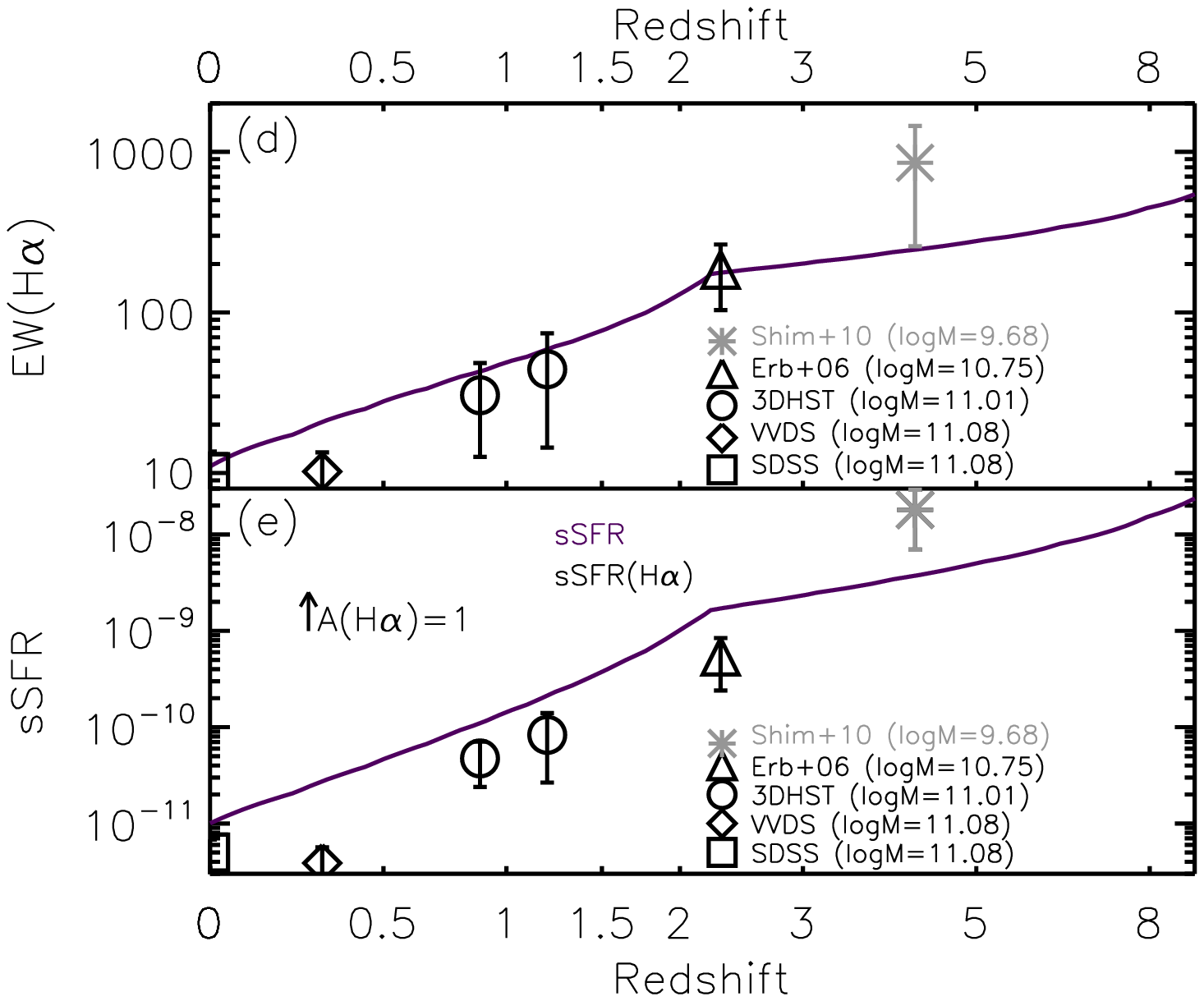} \\

\caption{Comparison of observed EW(H$\alpha$) with predictions from a simple observational supported model, at different redshifts. (a) Input SFH, and H$\alpha$ Luminosity (b) Luminosities at 6563 $\rm\AA$, from the Bruzual \& Charlot 2003 code. (c) Mass growth. (d) Evolution of EW(H$\alpha$) with redshift. (e) Evolution of sSFR with redshift. Data points are mean EW(H$\alpha$)/sSFR(H$\alpha$) of observed galaxies with mass in a 0.3 dex bin around the typical mass of the model at a given redshift.}
\label{EW_BC03}
\end{figure*}

We compare the observed evolution of EW(H$\alpha$) to what
might be expected from other observations. 
We construct the typical SFH of a galaxy with
mass $\sim \rm 10^{11} M_{\odot}$ at z=0.
As a starting point, we assume that the cumulative
number density remains constant with redshift 
(similar to van Dokkum et al. 2010, Papovich et al. 2011, Patel et al. 2012).
We use the mass functions of Marchesini et al. (2009) and Papovich et al. (2011), and we show the resulting mass evolution in Figure \ref{EW_BC03}c.
We determine the SFR at these masses from Damen et al. (2009),
Papovich et al. (2011) and Smit et al. (2012), and we fit a simple
curve to these values (indicated
by the curve in Figure \ref{EW_BC03}a). This evolutionary history reproduces the
mass evolution well (Figure \ref{EW_BC03}c).

Next we calculate the implied EW(H$\alpha$):
L(H$\alpha$) is derived using Equation 2, and Bruzual \& Charlot 2003 models are
used to calculate the stellar continuum, assuming solar metallicity (Figure \ref{EW_BC03}b).
The predicted EW(H$\alpha$) rises monotonically to high redshift, reaching
400$\rm\AA$  at $z$=8 (Figure \ref{EW_BC03}d). 
The predicted EW(H$\alpha$) corresponds surprisingly well to the observed EW(H$\alpha$) 
Even the $z$=4 detections by Shim et al. (2011) based on broadband IRAC photometry
are consistent within the errors. Apparently, our simple method produces
a robust prediction of the evolution of EW(H$\alpha$).
We note that the implied sSFR (Figure \ref{EW_BC03}e) is higher than expected from straight
measurement of L(H$\alpha$), consistently with significant dust
extinction. One magnitude of
extinction for H$\alpha$ is needed to reconcile this discrepancy.

It is remarkable that our prediction worked well for EW(H$\alpha$):
the average evolution of galaxies was derived from the evolution of the
mass function and SFR, which carry significant
(systematic) uncertainties when derived from observations; whereas the
EW(H$\alpha$) is a direct observable.

\section{Conclusions}
We have used the 3D-HST survey to measure the evolution of the EW(H$\alpha$) from z=0 to z=2.
We show that the EW(H$\alpha$) evolves strongly
with redshift, at a constant mass, like $(1+z)^{1.8}$. The evolution is independent
of stellar mass. The equivalent width goes down with mass (at constant redshift).
The increase with redshift demonstrates the strong evolution of star forming galaxies, using a consistent and completely model independent indicator.
We explore briefly the implied sSFR evolution, ignoring dust extinction. We find that the evolution with redshift is strong (sSFR $\sim (1+z)^{3.2}$). This  stronger evolution is expected as the mass-to-light ratio of galaxies evolves with time, and this enters the correction from EW to
sSFR. The increase with redshift is faster that predicted by semi-analytical
models (e.g., Guo \& White 2008), consistent with earlier results.

We construct the characteristic SFH of a $\rm 10^{11} M_{\odot}$ galaxy.
This simple history reproduces the observed evolution of the EW(H$\alpha$)
to $z$=2.5, and even to $z$=4. It implies that the EW(H$\alpha$) continue to increase to higher redshifts, up to 400 $\rm\AA$ at $z$=8. This has a significant impact for the photometry and spectroscopy of these high redshift sources. 

The study can be expanded in the future when the entire 3D-HST survey will be available, doubling the sample and including the ACS grism. In addition to increased statistics, the ACS grism will allow evaluation of the Balmer decrement and therefore a precise dust corrected evaluation of SFR. Moreover, a statistically significant H$\alpha$ sample at $z\sim 1$ will be central to understand the composition, the scatter and the physical origin of the so called 'star-forming-main sequence'.

\acknowledgments 
We thank the referee for providing valuable comments, and Jarle Brinchmann, David Sobral and Simone Weinmann for useful discussions. 
We acknowledge funding from ERC grant HIGHZ no. 227749.

\begin{table*}[!h!]
  \caption{Slopes of EW-$z$ and sSFR-$z$ Relation}
  \label{Slopes}
  \begin{center}
    \leavevmode
\begin{tabular}{lllll} \hline \hline            
$\rm log(M_{*}/M_{\odot})$    &       EW(det)   &   EW(all)  &    sSFR(det)    &  sSFR(all)      \\ \hline 
10.0-10.5& $ 1.79\pm  0.18$ & $ 1.52\pm  0.21$ & $ 3.32\pm  0.08 $ & $ 3.06\pm 0.13 $ \\ 
10.5-11.0& $ 1.89\pm  0.20$ & $ 1.75\pm  0.13$ & $ 3.18\pm  0.09 $ & $ 3.11\pm 0.18 $ \\ 
11.0-11.5& $ 2.21\pm  0.22$ & $ 2.12\pm  0.43$ & $ 3.50\pm  0.12 $ & $ 3.45\pm 0.26 $ \\ 
\hline
\vspace{0.2cm}
\end{tabular}

  \end{center}
\end{table*}

\end{document}